# Universal 3:1 Scaling of Quantum-Confined Stark Spectra Revealed by a Three-Dimensional Profile

Sha Han[1], Kebei Chen[1], Runnan Zhang[1], Juemin Yi[1], Wentao Song*[1] and Ke Xu*[1,2,3]

[1)] Platform for Characterization and Test, Suzhou Institute of Nano-Tech and Nano-Bionics, Chinese Academy of Sciences (CAS), Suzhou 215123, Jiangsu, People's Republic of China

[2)] Suzhou Nanowin Science and Technology Co, Ltd., Suzhou, 215123, Jiangsu, China

[3)] Jiangsu Institute of Advanced Semiconductors, Suzhou 215125, China

Corresponding authors: wtsong2017@sinano.ac.cn and kxu2006@sinano.ac.cn

**Abstract** We report that the quantum-confined Stark effect spectrum exhibits a nearly rigid redshift while preserving its characteristic peak spacing patterns when increasing the electric field strength $F$. Using InGaN as a model system, we uncover two electric-field-independent scaling laws for the spectral peaks in both the sub-bandgap and above-bandgap regions and the coefficient ratio is near 3:1. With a novel three-dimensional (3D) visualization, we reveal that the sub-bandgap peak spacings scale as $\frac{12\pi\hbar^2}{L^2\sqrt{m_e m_h}}$ while the above-bandgap peak spacings scale as $\frac{4\pi\hbar^2}{L^2\sqrt{m_e m_h}}$, explaining the origin of the 3:1 ratio. This scaling behavior, validated in both InGaN and GaAs systems and at electroluminescence working conditions, shows that increasing $F$ only expands the energy range and increases the number of peaks without altering the spacing. Beyond these laws, the 3D profile offers new insights into the Tauc background, Franz–Keldysh oscillations and coherence length, providing a powerful tool for the design and diagnostics of electro-optic devices.

***Introduction—***In quantum wells, scaling laws govern a multitude of properties from exciton binding energies [1-5], band gap [6], intersubband transition energy [7], dielectric properties [8] to carrier concentrations [9-15] and photoconductivity [16-18]. Among these, the quantum-confined Stark effect (QCSE) is of paramount importance, bridging fundamental quantum confinement physics with critical electro-optic phenomena [19–22]. However, conventional theoretical frameworks have primarily succeeded only in capturing isolated features of the QCSE spectrum: the field- and well-width-dependent fundamental energy levels $E_m$ [23–29], the red-shifted absorption edge $E_{11}$ [24,26-40] and specific higher-order transitions $E_{nn}$ ( $n > 1$ ) [25-27,32-34,38-45].

This absence of a global perspective hinders a comprehensive understanding: Variational methods [23-27,30,31,46] are typically restricted to the dominant low-energy transition $E_{11}$, struggling with the dense high-energy spectrum. While powerful, one-dimensional envelope function approximations [23-38,41-43], multi-band effective mass theory [27,28,32,33,38,40,43,44] and tight-binding models [27,37.39,40] often focus on individual transitions $E_{nn}$, lacking a holistic visualization that captures the collective behavior of all possible interband transitions under an electric field. Consequently, these methods fail to reveal the underlying scaling laws governing the average peak spacing or to establish an intuitive connection between spectral features and the quantum-confined electronic structure. This absence of a global perspective hinders a comprehensive understanding of the QCSE spectrum as a whole.

Here, we overcome these limitations by introducing a novel 3D profile of the interband transitions. We first observe that the QCSE spectrum undergoes a nearly rigid redshift with increasing electric field strength $F$ while preserving its characteristic peak spacing patterns. Through simulations, we then uncover two electric-field-independent scaling laws for the average peak spacing in both sub-bandgap and above-bandgap regions, following a $1/L^2$ dependence. We introduce a 3D profile which not only explains the origin of these universal scaling laws but also provides new insights into well-known phenomena such as the Tauc background, Franz-Keldysh oscillations (FKOs) and coherence length, offering a unified framework for understanding a range of electro-optic phenomena. These findings provide a powerful tool for the rational design and diagnostics of electro-optic devices.

***Simulations—***The electronic structure is simulated by solving the Schrödinger equation (Eq.1) for carriers in a triangular potential well of depth $eFL$, representing the quantum-confined Stark effect in a single quantum well (Fig. 1a). Interband transition probabilities between these confined states are computed using Eq.2 and Eq.3 with the key output being the average energy spacing between adjacent peaks in the simulated absorption spectrum for defined sub-bandgap and above-bandgap regions.

The primary material system is $In_{0.1}Ga_{0.9}N$(InGaN) with parameters: heavy-hole mass $m_{hh} = 2.00m_0$, electron mass $m_e = 0.18m_0$, bandgap $E_g = 3.0$ eV [47] and vacuum energy $E_{vac} = 5$ eV [48]. The default number of states evaluated is 100 calculated from $N_h = \frac{2}{3\pi}\frac{\sqrt{2m_h eFL^3}}{\hbar}$ [49] (increased to 200 for Fig.5). We explore a parameter space relevant to real devices: well width $L$ ranges from 5 to 12 nm and electric field $F$ from 1.0 to 3.0 MV/cm, spanning typical InGaN LED operating conditions up to the GaN breakdown field [50,51].

To validate the generality of our findings, comparative simulations are performed for GaAs quantum wells using GaAs parameters ( $m_e = 0.06m_0$ , $m_{hh} = 0.50m_0$ , $E_g = 1.424$ eV , $F \approx 0.5$ MV/cm) [52,53]. The well width $L$ is adjusted so the number of bound states is comparable to the InGaN reference case (see SM.2). To demonstrate universality in electroluminescence, InGaN is also simulated at three Fermi energies (0.1 eV, 0.2 eV, and 0.3 eV above the conduction band bottom), covering typical operational conditions [54-57]. The vacuum thickness is 10 nm on both sides, with a simulation step size of 0.05 nm.

$$\left(-\frac{\hbar^2}{2m_i}\frac{d^2}{dz_i} \pm eFz_i\right)\phi_{in}(z_i) = E_{in}\phi_{in}(z_i) \quad \text{(Eq.1)}$$

$$I_{nm} = \frac{\left|\int_0^L \phi_{e,n}(z)\phi_{h,m}(z)dz\right|^2}{\int_0^L |\phi_{e,n}|^2 dz \int_0^L |\phi_{h,m}|^2 dz} \quad \text{(Eq.2)}$$

$$\alpha \propto \sum_{n,m} I_{nm}\Theta\left(E - E_g - E_{e,n} - E_{h,m}\right) \quad \text{(Eq.3)}$$

***Results and Discussions—***

**A. Nearly rigid spectral redshift and the hypothesis of intrinsic spacing**

The QCSE model is shown in Fig.1(a) with the quantum states visualized. The QCSE spectrum shows a nearly rigid redshift for both sub-bandgap region and above-bandgap region while the characteristic peak spacing patterns remains consistent with increasing $F$ with InGaN as the example (Fig.1(b)). As $F$ increases, the entire QCSE spectrum shifts almost rigidly to lower energies while preserving its peak spacing pattern. The dominant optical transition may vary—for instance, shifting from the transition '17' (between the 1st electronic state and the 7th hole state) at 2.0 MV/cm to transition '31' at 2.2 MV/cm. A similar redshift is also discernable in previous experiments too (see e.g. Fig.1 in [25]). In addition, the observed nearly rigid redshift with increasing $F$ indicates that more spectral peaks are pushed into the sub-bandgap region. Notably, above-bandgap peaks also exhibit a nearly rigid redshift, though the corresponding peak spacing becomes narrower than that in the sub-bandgap region. Overall, the nearly rigid spectral redshift suggests that the average peak spacing is only weakly dependent on $F$ while the increasing field primarily results in a larger number of transitions appearing within both the sub-bandgap region and above-bandgap region. This observation suggests a remarkable possibility: the characteristic energy spacing between peaks may be an intrinsic property of the quantum well geometry, largely independent of the applied electric field strength $F$. If true, it would imply the existence of universal scaling laws governing the spectral density of the QCSE.

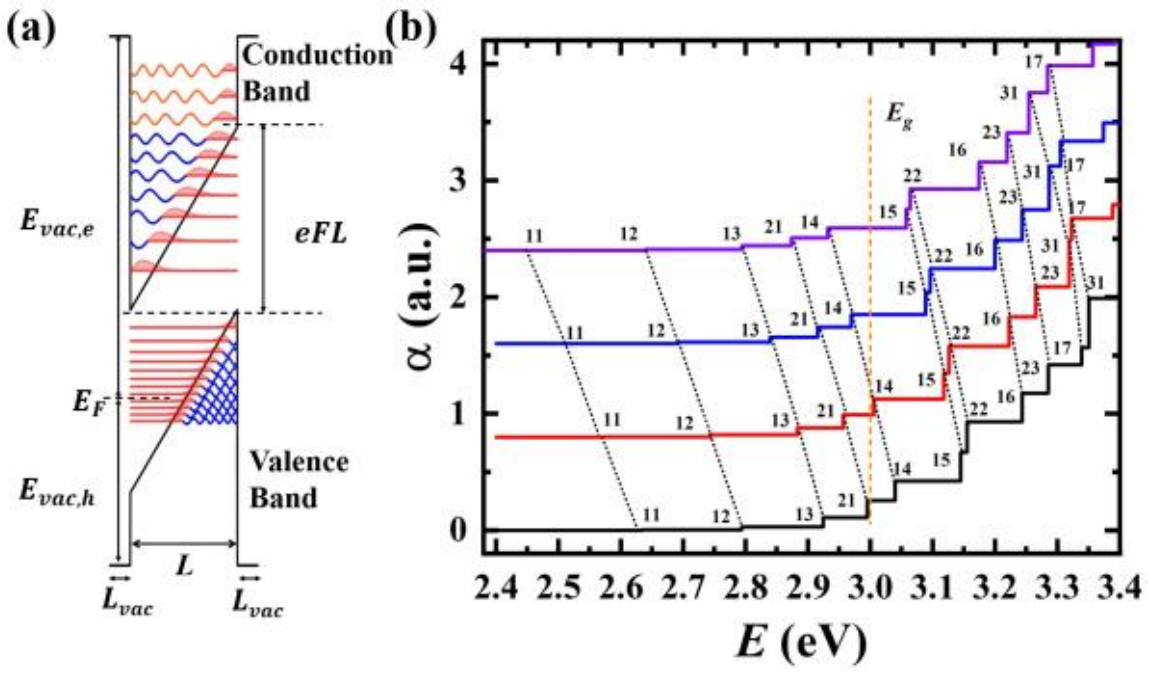

FIG. 1. (a) Schematic of the simulation model for an InGaN quantum well under an electric field. A triangular potential well of depth $eFL$ is formed for both electrons (top) and holes (bottom). Representative eigenfunctions within (blue loops with pink tails) and outside (orange loops) the well are shown. (b) Nearly rigid redshift of the QCSE spectrum for a well width $L = 5$ nm with increasing $F$ (2.0 MV/cm (black), 2.2 MV/cm (red), 2.4 MV/cm (blue) and 2.6 MV/cm (purple). Labels '12' denote transitions between the 1st electron state and the 2nd hole state. Black dashed lines guide the eye for the redshift.

### B. Discovery of universal $1/L^2$ scaling laws

To test this hypothesis, we systematically simulate the QCSE spectrum for varying values of $L$ and $F$ and calculate the average peak spacing in both the sub-bandgap region ($\overline{\Delta E_{tail}}$) and the above-bandgap region ($\overline{\Delta E_{FKO}}$). The above-bandgap region is defined as the energy range from $E_g$ to $2.05E_0$ where $E_0 = \left(\frac{e^2F^2\hbar^2}{2\mu}\right)^{1/3}$ is the characteristic Franz-Keldysh energy, $\mu$ is the reduced mass and $2.05\ E_0$ corresponds to the energy of the first FKO from $n\pi = \frac{4}{3}\left(\frac{E_n - E_g}{E_0}\right)^{3/2} - \frac{\pi}{4}$ [58-62]. As shown in Figure 2(a), we simulate the behavior of $\overline{\Delta E_{tail}}$ for the sub-bandgap region and $\overline{\Delta E_{FKO}}$ for the above-bandgap region as a function of $F$ from 1.0 MV/cm to 3.0 MV/cm. This range is based on the typical operational electric field of InGaN LED (1.5 MV/cm) [50] and the breakdown electric field of GaN PN diode (3.0 MV/cm) [51]. Both the $\overline{\Delta E_{tail}}$ and the $\overline{\Delta E_{FKO}}$ exhibit minor fluctuations with increasing $F$ and these fluctuations diminish as $L$ increases from 6 nm to 10 nm. Additionally, both the $\overline{\Delta E_{tail}}$ and the $\overline{\Delta E_{FKO}}$ decrease with the increasing $L$ (from square to triangle marker) and $\overline{\Delta E_{tail}}$ is consistently larger than the $\overline{\Delta E_{FKO}}$ under identical conditions.

More importantly, both spacings exhibit a strong and systematic dependence on the well width $L$ as shown in Fig. 2(b) that both trends follow an approximate $1/L^2$ scaling law with fitted coefficient of 3.172 eVnm$^2$ for $\overline{\Delta E_{tail}}$ and 1.231 eVnm$^2$ for $\overline{\Delta E_{FKO}}$. The nearly 3:1 ratio exist for GaAs too as shown in Fig.S1 that the coefficient is 11.052 eV·nm² for $\overline{\Delta E_{tail}}$ and 4.500 eV·nm² for $\overline{\Delta E_{FKO}}$. Further, both $\overline{\Delta E_{tail}}$ and $\overline{\Delta E_{FKO}}$ exhibit significant error bars which correspond to closely spaced spectral peaks in the simulations. The error bar of $\overline{\Delta E_{tail}}$ is comparatively larger, a point that will also be addressed in section D.

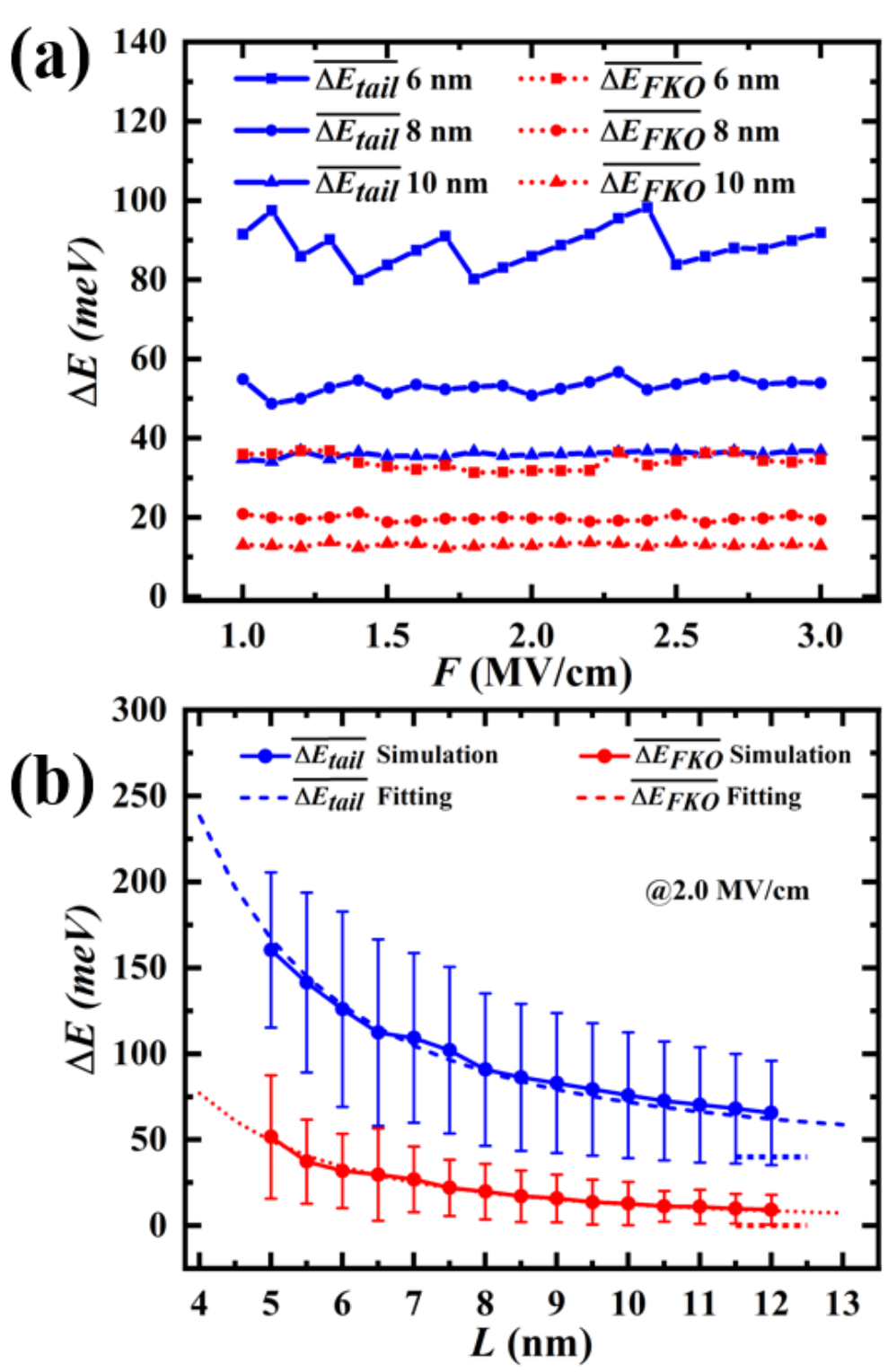

FIG. 2. Scaling of average peak spacing with electric field and well width. (a) Average peak spacing in the sub-bandgap ($\overline{\Delta E_{tail}}$) and above-bandgap ($\overline{\Delta E_{FKO}}$) regions as a function of $F$ for different well widths $L$. Error bars represent the standard deviation of peak spacings within the defined energy windows. (b) The scaling laws of $\overline{\Delta E_{tail}}$ and $\overline{\Delta E_{FKO}}$ with $L$ at $F =$ 2.0 MV/cm. Solid lines: simulation data. Dashed lines: fits with coefficients of 3.172 eVnm$^2$ (blue) and 1.231 eVnm$^2$ (red).

### C. Geometric origin of the scaling laws: A 3D transition profile

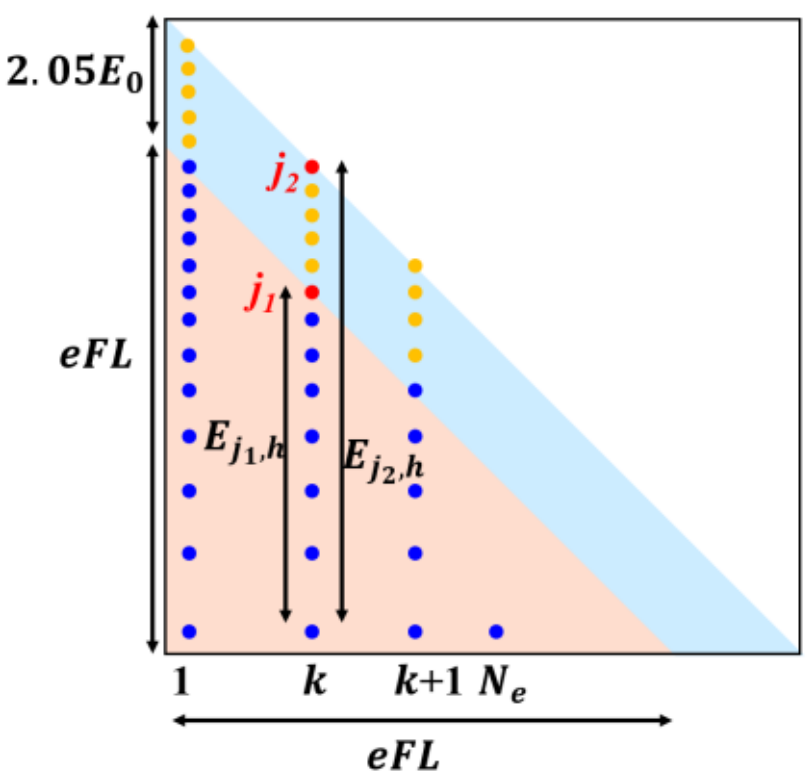

FIG. 3. Schematic of the 3D profile for interband transitions. Electron energy $E_e$ and hole energy $E_h$ form the axes. The transition strength is represented along the z-axis (not shown). The pink-shaded triangle marks the sub-bandgap region (transition energy $E < E_g$, bounded by the line $E_e + E_h = E_g$). The blue-shaded region marks the above-bandgap window up to the first FKO maximum. For a given electron state $k$, vertical cuts intersect these boundaries at hole indices $j_1(k)$ and $j_2(k)$, defining the number of contributing transitions in each region.

What is the physical origin of this universal $1/L^2$ scaling and the fixed 3:1 ratio? To answer this, we introduce a novel 3D profile to analyze the interband transitions (Fig. 3). The core physical idea is to map the entire set of possible transitions onto a 2D grid defined by electron energy $E_e$ and hole energy $E_h$. In this space, lines of constant transition energy are diagonal. The key insight is that the number of allowed transitions within a given energy window can be calculated by counting states in this $(E_e, E_h)$ space. We employ the Wentzel-Kramers-Brillouin (WKB) approximation [63] to describe the bound states in the triangular wells. Assuming a sufficiently dense spectrum, the discrete sums over states can be approximated by continuous integrals. This powerful approach allows us to derive analytical expressions for the total number of peaks in the sub-bandgap and above-bandgap regions (see SM.2 for detailed derivations). The central finding is that the total number of sub-bandgap transitions scales as $\sim FL^3$ while the number of above-bandgap transitions scales as $\sim F^{\frac{2}{3}}L^2$ (SM.2). Interestingly, the energy widths of these regions also scale differently with $F$: the sub-bandgap region spans an energy $eFL$ while the above-bandgap region spans an energy $2.05E_0 \sim F^{\frac{2}{3}}L^2$ (SM.2).

The average peak spacing is given by the ratio of the energy width to the number of peaks. This division cancels the dependence on $F$, leaving a pure ${}^{1}/_{L^2}$ scaling for both regions: $\overline{\Delta E_{tail}} = \frac{12\pi\hbar^2}{L^2\sqrt{m_e m_h}}$ and $\overline{\Delta E_{FKO}} = \frac{4\pi\hbar^2}{L^2\sqrt{m_e m_h}}$ (see SM.2). The striking 3:1 ratio is a direct manifestation of the QCSE geometry, indicating that the above-bandgap peaks distribute more densely than the sub-bandgap peaks. Next is the discussion of $F$'s inert dispersion. As shown in Figure 2(a), the trend with $F$ has small fluctuations for 6 nm (squares). The fluctuations stem from the relatively few peak numbers for smaller $L$ since the sub-bandgap region has $0.0217 F_u L_u^3$ peaks and the above-bandgap region has $0.0847 F_u^{\frac{2}{3}} L_u^2$ peaks (from $N_i$ in SM.2) with $F_u$ the electric field strength in unit of MV/cm and $L_u$ the quantum well width in unit of nm. For larger $L$, the peak numbers increase rapidly. Furthermore, the relatively minor influence of $F$ on the peak spacings has been observed in GaAs quantum wells by the electroreflectance [64] and the photocurrent [25]. Besides, previous relative constant behavior of single $E_{nm}$ with $F$ for small $L$ (10nm) and dispersive behavior of $E_{nm}$ with $F$ for large $L$ (20nm) in GaAs quantum wells could be understood [27] since for small $L$ the potential $eFL$ is weak and $F$ could be treated as zero while for large $L$ the influence of $F$ can no longer be ignored.

It is evident that despite the approximations employed, the analytical results derived from 3D profile could capture the main scaling behavior observed in the simulations reasonably well (scatters) with the deduced coefficient (4.393 eVnm$^2$) for the $\overline{\Delta E_{tail}}$ exceeds the simulated coefficient (3.172 eVnm$^2$) by approximately 38% and the deduced coefficient (1.464 eVnm$^2$) for the $\overline{\Delta E_{FKO}}$ is about 19% larger than the simulated coefficient (1.231 eVnm$^2$). The observed ${}^{1}/_{L^2}$ scaling law also coincides with the previous tight-binding result that the allowed transition $E_{nn}$ scales with $L$ in a cosine function since the first term in the cosine function is a $^{1}/_{L^2}$ term [39,40].

## D. Universality of the scaling laws and model refinements

The universality of these derived scaling laws is further confirmed by applying our model to GaAs quantum wells. As shown in the Fig. S1 in the SM.1, the simulated average peak spacings for GaAs also follow the predicted $^{1}/_{L^2}$ dependence with coefficients that agree well with the theoretical prediction $\overline{\Delta E_{tail}} = \frac{12\pi\hbar^2}{L^2\sqrt{m_e m_h}}$ and $\overline{\Delta E_{FKO}} = \frac{4\pi\hbar^2}{L^2\sqrt{m_e m_h}}$ when using GaAs-specific effective masses. This successful application to a materially distinct system (GaAs vs. InGaN) with different effective masses and characteristic fields strongly validates that the scaling laws are universal, governed solely by the quantum well width and the carrier effective masses.

We further demonstrate the universality of the scaling laws against variations in the Fermi level, confirming their applicability under realistic device operating conditions, for e.g. electroluminescence working conditions. The inclusion of the $E_F$ would bend the conduction bottom into a curve as shown in Fig.4(a) which violates the criteria of the 3D profile. The curve would bend more with the increase of $E_F$. However, as illustrated in Fig. 4(b) and 4(c), for typical $E_F$ nearly all the data collapse to the results with zero $E_F$, indicating the insensitivity of the scaling laws to moderate variations in carrier concentration. Here, we consider three representative values of $E_F$ (1.60 eV, 1.70 eV, and 1.80 eV above midgap, corresponding to 0.10–0.30 eV above the conduction band bottom), covering typical ranges encountered in electroluminescence devices [54-57]. The validation via experiment is possible since the GaN PIN shows clear steps at room temperature [65]. The fact that the same 3:1 scaling, governed solely by well width and effective masses, emerges across materially distinct systems (III-nitrides vs. III-arsenides) and under operational conditions (with finite carrier density) underscores its fundamental, universal nature.

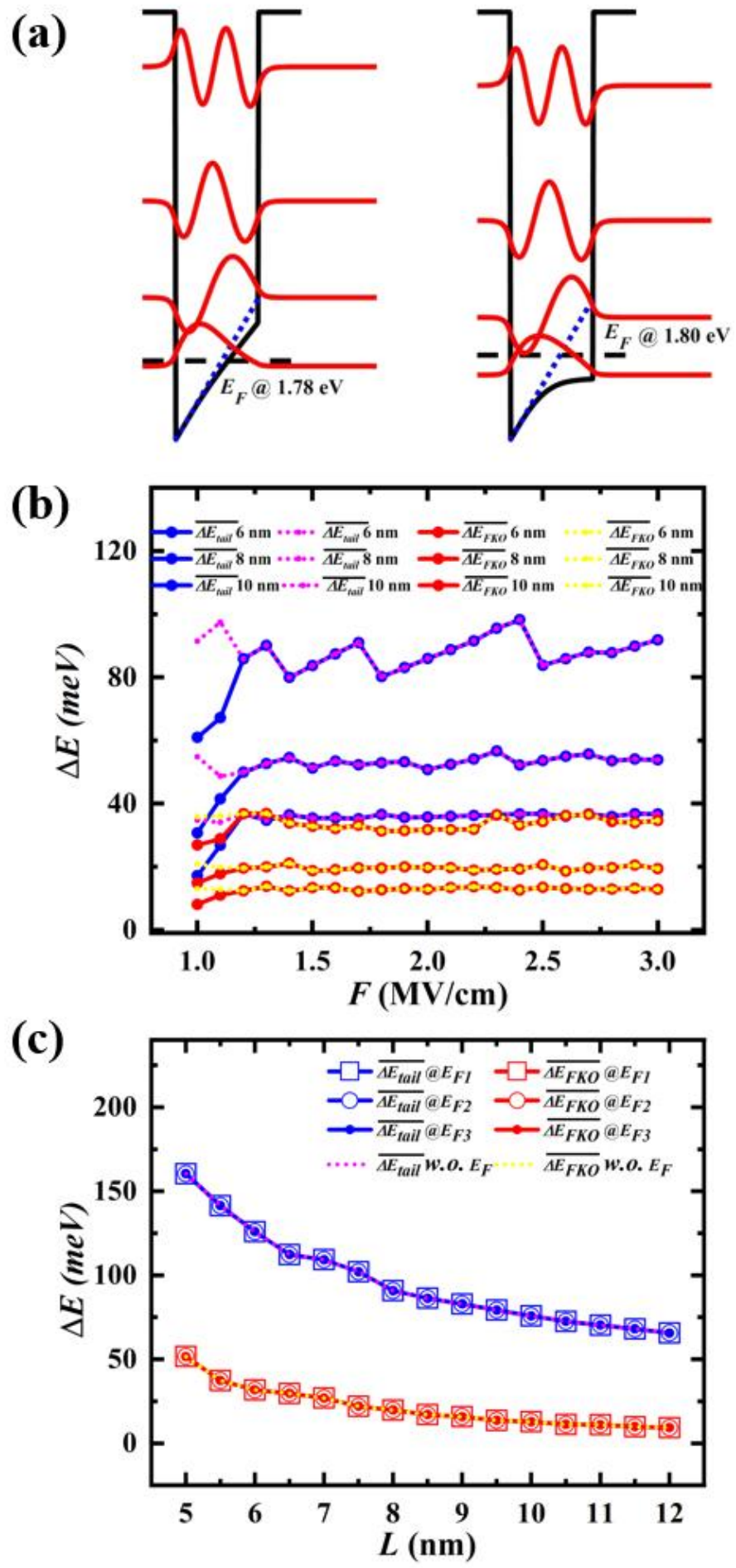

FIG. 4. (a) Band structure modification under Fermi level shift. Conduction band profile at Fermi energies of 1.78 eV (left) and 1.80 eV (right) above mid-gap for $L$ = 5 nm and $F$ = 1.0 MV/cm. The original band edge (blue dotted) is modified by carrier screening (black solid). (b) Average peak spacing vs. $F$ for $L$ = 6, 8, 10 nm at the highest Fermi energy $E_{F3}$ (1.8eV). The results for 0 eV $E_F$ are shown by the dotted lines (pink for $\overline{\Delta E_{tail}}$ and yellow for $\overline{\Delta E_{FKO}}$ ). (c) Scaling with $L$ at $F$ = 2.0 MV/cm for three different Fermi energies. Line styles, fitting curves and theoretical curves are identical to those in Fig. 2(b). The results for 0 eV $E_F$ are shown by the dotted lines (pink for $\overline{\Delta E_{tail}}$ and yellow for $\overline{\Delta E_{FKO}}$).

The estimation of the origins for deviation of the theoretical scaling laws is shown in SM.3. For $\overline{\Delta E_{tail}}$ (Fig.S2(a)), 3 factors, the overestimation of the sub-bandgap energy span as $eFL$, the repeat counting of the point $j_1$ and the omission term $0.25N_e$ in Eq. 6 in SM.2 are balanced to provide a better estimation for the simulation. For $\overline{\Delta E_{FKO}}$ (Fig.S2(b)), 2 factors, the omission term $N_e$ and the repeat counting of both point $j_1$ and $j_2$, are concluded. The original deduction is good enough and the refined deduction with only the omission term $N_e$ provides an even better estimation.

## E. Unified interpretation of spectral reatures and diagnostic applications

Beyond deriving scaling laws, the 3D profile offers profound insights into well-established phenomena such as the Tauc-background and FKOs as illustrated in Figure 5. As demonstrated in Fig. S5(a1) in SM.6, the line with slope $\frac{m_e}{m_h}$ is the Tauc-background under zero field condition. Accordingly, the stripes with slope nearly $\frac{m_e}{m_h}$ in Figure 5(b) can be regarded as remnants of the Tauc-background. Unlike the zero-field case where transition probabilities are uniform (Fig. S5(a1)), the presence of an electric field modifies these transitions, resulting in higher probabilities at lower $E_e$ that gradually diminish as $E_e$ increases (Fig. S5(a2)). The 135° stripes in Figure 5(c) correspond to FKOs with their intensity decreasing from left to right. The FKOs appear at 135° stripes since the states in the triangle potential well are Airy-function-like ( $Z_{in} = -\left(\frac{2m_i}{q^2\hbar^2F^2}\right)^{\frac{1}{3}}(E_{in} \mp qFz_i)$ with $i$ for e or h) that the energy and the space are connected linearly. Therefore, increasing the electron state energy and decreasing the hole state energy by a same energy means shifting both the electron and hole state by a same distance, resulting nearly the same transition intensity compared with the intensity before the shift. Thus the intensity along the 135° line are at the same level. Additionally, the interplay between the Tauc-background-like features and the FKOs produces a checkerboard pattern in the 3D profile. Furthermore, the 3D profile provides a natural explanation for the observed similarity between the quantum well width $L$ and the coherence length $L_{ch}$ reported in Refs. [66-69]. As illustrated in the SM.7 Fig. S6, the geometric relation $\frac{L_{ch}}{L} < \frac{BE}{OA} = \frac{1}{\sqrt{2}}\frac{m_h+m_e}{m_h-m_e}$ emerges naturally from the profile, explaining why $L$ and $L_{ch}$ are often comparable.

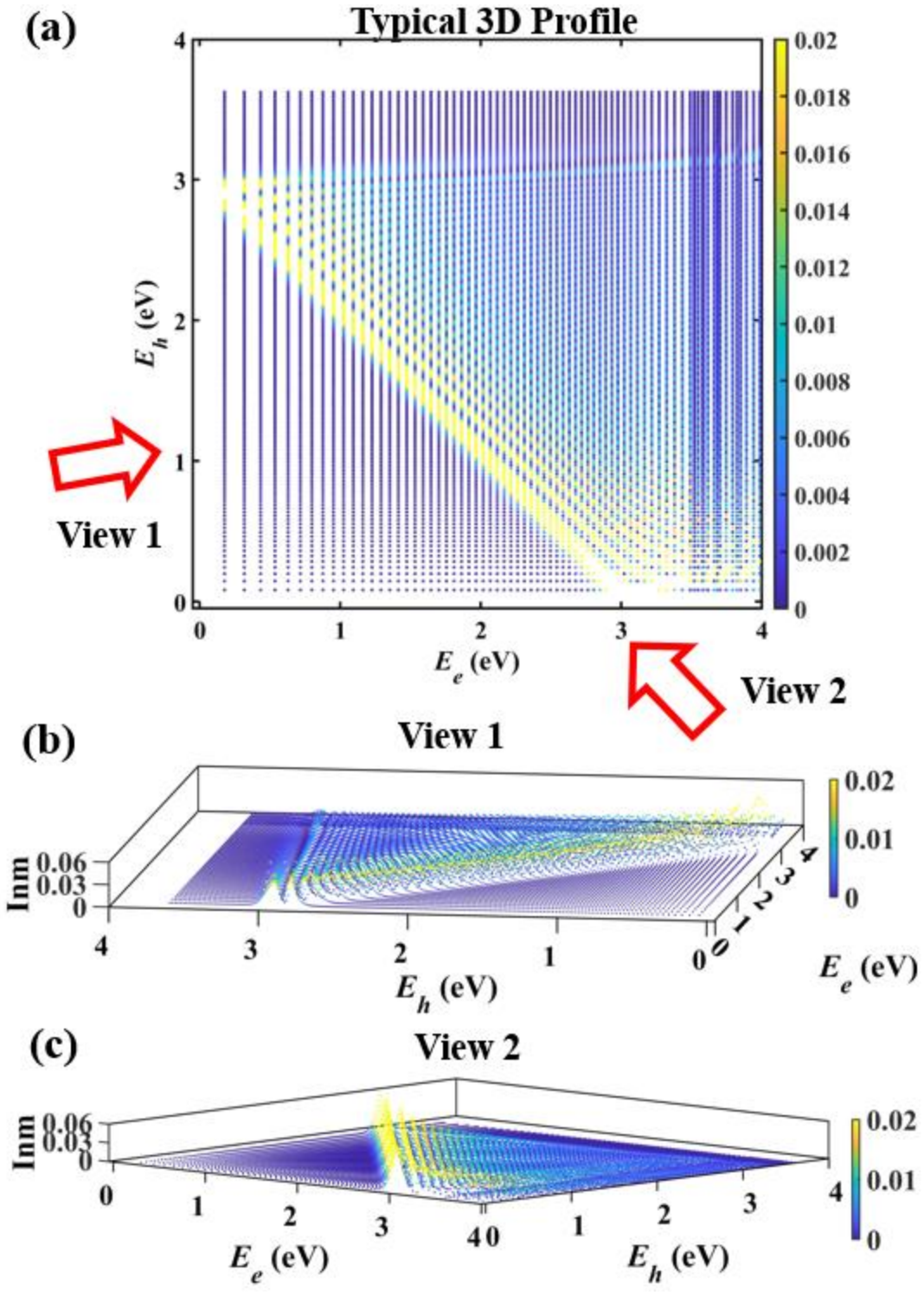

FIG. 5. Simulated 3D profile and its interpretation for a 60 nm well at $F = 0.5$ MV/cm. (a) Full 3D profile with color bar denotes the transition probability. (b) View 1: Stripes with slope near $m_e/m_h$ (red dashed line) are identified as remnants of the Tauc background. (c) View 2: Stripes along 135° (constant transition energy) correspond to FKOs. Valleys between these stripes coincide with FKO extrema in the absorption spectrum.

Furthermore, the 3D profile establishes a powerful spectral map for device design and diagnostics. The smaller energy spacings in GaN compared to GaAs indicate superior monochromaticity in GaN-based optoelectronic devices. In addition, the narrower energy span observed in materials with smaller $F$ further confirms that non-polar devices exhibit better monochromatic performance and would have a blueshift for the emission compared with the polar devices. The 3D profile could also provide the type of the electric field and the value of the field strength $F$ (Fig.S5 in SM.6). By comparing an experimentally constructed profile—derived from scanning tunneling spectroscopy [70] (for eigenenergies) and photocurrent spectroscopy [25] (for transition intensities)—with theoretical predictions, one can identify spectral anomalies. Unanticipated transitions, deviations from the characteristic 135 ° Franz–Keldysh oscillation stripes, or a disrupted Tauc-background pattern may signal the presence of defects, unintended doping, or undesired internal fields. Besides, the 3D profile for quantum wells could also indicate the band alignment type. Rooted directly in the electronic structure, this diagnostic capability offers a principled approach to pinpoint failure mechanisms in electro-optic devices, thereby complementing conventional characterization methods.

***Conclusions***—In conclusion, we have discovered two universal scaling laws that govern the quantum-confined Stark effect spectrum. Arising from the observation of a nearly rigid redshift, the average peak spacing below and above the bandgap scales as $\frac{12\pi\hbar^2}{L^2\sqrt{m_e m_h}}$ and $\frac{4\pi\hbar^2}{L^2\sqrt{m_e m_h}}$, respectively, maintaining a 3:1 ratio independent of the applied electric field. These laws, validated across material systems (InGaN, GaAs) and under operating conditions, are explained and unified by a novel 3D profile of interband transitions. This geometric framework not only reveals the origin of the scaling but also seamlessly interprets disparate spectral features (Tauc background, FKOs) and links the well width to the coherence length. Our work provides a new paradigm for analyzing electro-optic spectra—transforming them into intuitive, geometric maps for both fundamental understanding and device diagnostics.

*Acknowledgements*—One paragraph provides acknowledgement to financial support from National Natural Science Foundation of China (12204511 and 62474188), National Key Research and Development Program of China (2021YFA1201503) and CAS Pioneer Hundred Talents Program (W. Song) and the Fundamental Research Pilot Project of Suzhou (SJC2021007 and SJC2021010).

[1] D. Campi and C. Villavecchia, Excitonic properties in semiconductor quantum wells: Numerical calculations and scaling behavior, IEEE journal of quantum electronics **28**, 1765 (1992)

[2] J. Heckötter, M. Freitag, D. Fröhlich, M. Aβmann, M. Bayer, M. A. Semina and M. M. Glazov, Scaling laws of Rydberg excitons, Phys. Rev. B **96**, 125142 (2017).

[3] J. -C. Blancon, A. Stier, H. Tsai, W. Nie, C. Stoumpos, B. Traoré, L. Pedesseau, M. Lepenekian, S. Tretiak, S. Crooker *et al*., Unusual thickness dependence of exciton characteristics in 2D perovskite quantum wells, arXiv:1710.07653 (2017).

[4] X. Ahmad, M. Zubair, O. Jalil, M. Q. Mehmood, U. Younis, X. Liu, K. W. Ang. and L. K. Ang, Generalized scaling law for exciton binding energy in two-dimensional materials, Phys. Rev. Appl. **13**, 064062 (2020).

[5] J. -C. Blancon, A. V. Stier, H. Tsai, W. Nie, C. C. Stoumos, B. Traore, L. Pedesseau, M. Kepenekian, F. Katsutani, G. Noe *et al*., Scaling law for excitons in 2D perovskite quantum wells, Nat. Com. **9**, 2254 (2018).

[6] S. Eichfeld, C. Eichfeld, Y. Lin, L. Hossain and Joshua Robinson, Rapid, non-destructive evaluation of ultrathin $WSe_2$ using spectroscopic ellipsometry, APL Mat. **2**, 092508 (2014).

[7] C. Lim, M. Beeler, A. Ajay, J. Lähnemann, E. Bellet-Amalric, C. Bougerol and E. Monroy, Intersubband transitions in nonpolar GaN/Al(Ga)N heterostructure in the short- and mid-wavelength infrared regions, J. Appl. Phys. **118**, 014309 (2015).

[8] H. Gu, B. Song, M. Fang, Y. Hong, X. Chen, H. Jiang, W. Ren and S. Liu, Layer-dependent dielectric and optical properties of centimeter-scale 2D $WSe_2$: evolution from a single layer to few layers, Nanoscale **11**, 22762 (2019).

[9] N. K. Dutta, Calculated threshold current of GaAs quantum well lasers, J. Appl. Phys. **53**, 7211 (1982).

[10] P. Blood, E. D. Fletcher and K. Woodbridge, Dependence of threshold current on the number of wells in AlGaAs-GaAs quantum well lasers, Appl. Phys. Lett. **47**, 193 (1985).

[11] J. Nagle, S. Hersee, M. Krakowski, T. Weil and C. Weisbuch, Threshold current of single quantum well lasers: The role of the confining layers, Appl. Phys. Lett. **49**, 1325 (1986).

[12] A. Yariv, Scaling laws and minimum threshold currents for quantum-confined semiconductor lasers, Appl. Phys. Lett. **53**, 1033 (1988).

[13] M. P. Stopa and S. Das Sarma, Density scaling and optical properties of semiconductor parabolic and square quantum wells, Phys. Rev. B **45**, 8526 (1992).

[14] S. M. Shank, J. A. Varriano and G. W. Wicks, Single quantum well GaAs/AlGaAs separate confinement heterostructure lasers with *n*-type modulation doped cores, Appl. Phys. Lett. **61**, 2951 (1992).

[15] J. Gilor, I. Samid and D. Fekete, Threshold current density reduction of strained AlInGaAs quantum-well laser, IEEE J. Quan. Elec. **40**, 1355 (2004).

[16] R. Calarco, M. Marso, T. Richter, A. I. Aykanat, R. Meijers, A. v. d. Hart, T. Stoica and H. Lüth, Size-dependent photoconductivity in MBE-grown GaN-nanowires, Nano Lett. **5**, 981 (2005).

[17] H. Chen, R. Chen, F. Chang, L. Chen, K. Chen and Y. Yang, Size-dependent photoconductivity and dark conductivity of *m*-axial GaN nanowires with small critical diameter, Appl. Phys. Lett. **95**,

143123 (2009).

[18] H. Chen, R. Chen, N. Rajan, F. Chang, L. Chen, K. Chen, Y. Yang, M. Reed, Size-dependent persistent photocurrent and surface band bending in *m*-axial GaN nanowires, Phys. Rev. B **84**, 205443 (2011).

[19] D. A. B. Miller, D. S. Chemla, T. C. Damen, A. C. Gossard, W. Wiegmann, T.-H. Wood and C. A. Burrus, Band-edge electroabsorption in quantum well structures: the Quantum Confined Stark Effect, Phys. Rev. Lett. **53**, 2173 (1984).

[20] D. A. B. Miller, D. S. Chemla, T. C. Damen, A. C. Gossard, W. Wiegmann, T.-H. Wood and C. A. Burrus, Electric field dependence of optical absorption near the band gap of quantum well structures, Phys. Rev. B **32**, 1043 (1985).

[21] S. -C. Tsai, C. -H. Lu and C. -P. Liu, Piezoelectric effect on compensation of the quantum-confined Stark effect in InGaN/GaN multiple quantum wells based green light-emitting diodes, Nano. Ener. **28**, 373 (2016).

[22] S. Zhu, S. Lim, J. Li, Z. Yu, H. Cao, C. Yang , J. Li and L. Zhao, Influence of quantum confined Stark effect and carrier localization effect on modulation bandwidth for GaN-based LEDs, Appl. Phys. Lett. **111**, 171105 (2017).

[23] G. Bastard, E. E. Mendez, L. L. Chang and L. Esaki, Variational calculations on a quantum well in an electric field, Phys. Rev. B **28**, 3241 (1983).

[24] D. A. B. Miller, D. S. Chemla, T. C. Damen, A. C. Gossard, W. Wiegmann, T. H. Wood and C. A. Burrus, Band-edge electroabsorption in quantum well structures: the quantum confined Stark effect, Phys. Rev. Lett. **53**, 2173 (1984)

[25] K. Yamanaka, T. Fukunaga, N. Tsukada, K. Kobayashi and M. Ishii, Photocurrent spectroscopy in GaAs/AlGaAs multiple quantum wells under a high electric field perpendicular to the heterointerface, Appl. Phys. Lett. **48**, 840 (1986).

[26] M. Matsuura and T. Kamizato, Subbands and excitons in a quantum well in an electric field, Phys. Rev. B **33**, 8385 (1986).

[27] G. D. Sanders and K. K. Bajaj, Electronic properties and optical-absorption spectra of GaAs-$Al_xGa_{1-x}As$ quantum wells in externally applied electric fields, Phys. Rev. B **35**, 2308 (1987).

[28] K. Satzke, G. Weiser, W. Stolz and K. Ploog, Optical study of the electronic states of $In_{0.53}Ga_{0.47}As/In_{0.52}As_{0.48}As$ quantum wells in high electric fields, Phys. Rev. B **43**, 2263 (1991).

[29] T. Nakaoka, S. Kako and Y. Arakawa, Unconventional quantum-confined Stark effect in a single GaN quantum dot, Phys. Rev. B **73**, 121305 (2006).

[30] J. A. Brum and G. Bastard, Electric-field-induced dissociation of excitons in semiconductor quantum wells, Phys. Rev. B **31**, 3893 (1985).

[31] D. A. B. Miller, D. S. Chemla, T. C. Damen, A. C. Gossard, W. Wiegmann, T. H. Wood and C. A. Burrus, Electric field dependence of optical absorption near the band gap of quantum-well structures, Phys. Rev. B **32**, 1043 (1985).

[32] P. W. Yu, G. D. Sanders, K. R. Evans, D. C. Reynolds, K. K. Bajaj, C. E. Stutz and R. L. Jones, Electric field dependence of exciton transition energies in GaAs-$Al_xGa_{1-x}As$ quantum wells studied by photocurrent spectroscopy, Phys. Rev. **38**, 7796 (1988).

[33] Y. Kajikawa, M. Hata, N. Sugiyama and Y. Katayama, Photocurrent spectroscopy of a (001)- and a (111)- oriented $GaAs/Al_{0.33}Ga_{0.67}As$ quantum-well structure, Phys. Rev. B **42**, 9540 (1990).

[34] D. P. Wang, C. T. Chen, H. Kuan, S. C. Shei and Y. K. Su, Observation of quantum confined Stark effect in $In_xGa_{1-x}As/GaAs$ single-quantum well by photoreflectance spectroscopy, J. Appl. Phys. **78**, 2117 (1995).

[35] T. Takeuchi, C. Wetzel, S. Yamaguchi, H. Sakai, H. Amano and I. Akasaki, Determination of piezoelectric fields in strained GaInN quantum wells using the quantum-confined Stark effect, Appl. Phys. Lett. **73**, 1691 (1998).

[36] A. Jaeger and G. Weiser, Excitonic electroabsorption spectra and Franz-Keldysn effect of In0.53Ga0.47As/InP studied by small modulation of static fields, Phys. Rev. B **58**, 10674, (1998).

[37] Y. Kuo, Y. Lee, Y. Ge, S. Ren, J. E. Roth, T. I. Kamins, D. A. B. Miller and J. S. Harris, Strong quantum-confined Stark effect in germanium quantum-well structures on silicon, Nature **437**, 1334 (2005).

[38] K. Ryczko, G. Sęk, P. Sitarek, A. Mika, J. Misiewicz, F. Langer, S. Höfling, A. Forchel and M. Kamp, Verification of band offsets and electron effective masses in GaAsN/GaAs quantum wells: Spectroscopic experiment versus 10-band k • p modeling, J. Appl. Phys. **113**, 233508 (2013).

[39] G. Zhang, S. Huang, A. Chaves, C. Song, B. O. Özçelik, T. Low and H. Yan, Infrared fingerprints of few-layer black phosphorus, Nat. Com. **8**, 14071 (2017).

[40] G. Zhang, S. Huang, F. Wang and H. Yan, Layer-dependent electronic and optical properties of 2D black phosphorus: Fundamentals and engineering, Laser Phot. Rev. **2021**, 2000399 (2021).

[41] R. T. Collins, L. Viña, W. I. Wang, L. L. Chang, L. Esaki, K. v. Klitzing and K. Ploog, Phys. Rev. B, **36**, 1531 (1987).

[42] N. Kotera, K. Tanaka and H. Nakamura, Photocurrent spectroscopy of 5-nm-wide InGaAs/InAlAs quantum wells and quadratic dependence of optical transition energies on quantum numbers, J. Appl. Phys. **78**, 5168 (1995).

[43] K. Tanaka, N. Kotera and H. Nakamura, Electric field effect for eigen states in $In_{0.53}Ga_{0.47}As/In_{0.52}Al_{0.48}As$ multi-quantum wells using photocurrent spectroscopy, 16th IPRM. 2004 International Conference on Indium Phosphide and Related Materials p318-321 (2004).

[44] Y. Song, D. Chen, L. Wang, H. Li, G. Xi and Y. Jiang, Splitting of valance subbands in the wurtzite *c*-plane InGaN/GaN quantum well structure, Appl. Phys. Lett. **93**, 161910 (2008).

[45] R. K. Schaevitz, J. E. Roth, S. Ren, O. Fidaner and D. A. B.

Miller, Materials properties of Si-Ge/Ge quantum wells, IEEE Journal of selected topics in quantum electronics **14**, 1082 (2008).

[46] E. H. Li, K. S. Chan, B. L. Weiss and J. Micallef, Quantum-confined Stark effect in interdiffused AlGaAs/GaAs quantum well, Appl. Phys. Lett. **63**, 533 (1993).

[47] Gorczyca, J. Plesiewicz, L. Dmowski, T. Suski, N. E. Christensen, A. Svane, C. Gallinat, G. Koblmueller, and J. Speck, Electronic structure and effective masses of InN under pressure, J. Appl. Phys. **104**, 013704 (2008).

[48] T. Schultz, R. Schlesinger, J. Niederhausen, F. Henneberger, S. Sadofev, S. Blumstengel, A. Vollmer, F. Bussolotti, J. P. Yang, S. Kera, K. Parvez, N. Ueno, K. Müllen and N. Koch, "Tuning the work function of GaN with organic molecular acceptors," Phys. Rev. B **93**, 125309 (2016).

[49] S. Han, J. Yi, W. Song, K. Chen, S. Zheng, Y. Zhang and K. Xu, "Interband transition physics from the absorption edge in GaN: New prospects from numerical analysis," AIP Advances 13, 125016 (2023).

[50] H. Ryu, K. Jeon, M. Kang, H. Yuh, Y. Choi and J. Lee, A comparative study of efficiency droop and internal electric field for InGaN blue light-emitting diodes on silicon and sapphire substrates, Scientific reports 7, 44814 (2017).

[51] S. Mandal, M. Kanathila, C. Pynn, W. Li, J. Gao, T. Margalith, M. Maurent and S. Chowdhury, Observation and discussion of avalance electroluminescence in GaN PN diodes offering a breakdown electric field of 3 MV/cm, Semiconductor Science and Technology **33**, 065013 (2018).

[52] N. Bouarissa and H. Aourag, Effective masses of electrons and heavy holes in InAs, InSb, GaSb, GaAs and some of their ternary compounds, Infrared physics & techonology 40, 343 (1999).

[53] S. Tautz, S. Opel, P. Kiesel, S. U. Dankowski, H. M. Hauenstein, A. Seilmeier, M. Krause, U. D. Keif and G. H. Döhler, Spectrally and temporally resolved investigations of the carrier dynamics in biased low temperature grown GaAs, in European Quantum Electronics Conferenc (Optica Publishing Group, 1998) p. QMC4.

[54] A. Arif, H. Zhao, Y.-K. Ee, and N. Tansu, Spontaneous emission and characteristics of staggered InGaN quantum-well light-emitting diodes, IEEE Journal of Quantum Electronics **44**, 573 (2008).

[55] Y.K. Kuo, T.-H. Wang, J.-Y. Chang and M.-C. Tsai, Advantages of InGaN light-emitting diodes with GaN-InGaN-GaN barriers, Appl. Phys. Lett. **99**, 091107(2011).

[56] K. Bulashevich, O. Khokhlev, I. Y. Evstratov and S. Y. Karpov, Simulation of light-emitting diodes for new physics understanding and device design, in *Light-Emitting Diodes: Materials, Devices and Applications for Solid State Lighting XVI,* Vol. **8278** (SPIE, 2012) pp. 152-163.

[57] Z.-H. Zhang, Z. Kyaw, W. Liu, Y. Ji, L. Wang, S. T. Tan, X. W. Sun and H. V. Demir, A hole modulator for InGaN/GaN light-emitting diodes, Appl. Phys. Lett. **106**, 063501 (2015).

[58] J. Callaway, Optical absorption in an electric field, Phys. Rev.**130**, 549 (1963).

[59] K. Tharmalingam, Optical absorption in the presence of a uniform field, Phys. Rev. **130**, 2204 (1963).

[60] J. Callaway, Optical absorption in an electric field, Phys. Rev.**134**, A998 (1964).

[61] C. M. Penchina, Phonon-assisted optical absorption in an electric field, Phys. Rev. **138**, A924 (1965).

[62] D. E. Aspnes and A. A. Astudna, Schottky-barrier electroreflectance: Application to GaAs, Phys. Rev. B **7**, 4605 (1973).

[63] J. H. Davises, *The physics of low-dimensional semiconductor: an introduction* (Cambridge university press, 1998).

[64]J. Javaliauskas, G. Krivaite, A. Galickas, I. Šimklene, U. Olin and M. Ottosson, Quantum confined Stark effect in InGaAs/GaAs quantum wells under high electric fields, Phys. Stat. Sol. (b) **191**, 155 (1995).

[65] F. Zhang, M. Ikeda, K. Zhou, Z. Liu, J. Liu, S. Zhang and H. Yang, Injection current dependences of electroluminescence transition energy in InGaN/GaN multiple quantum wells light emitting diodes under pulsed current conditions. J. Appl. Phys. **118**, 033101 (2015).

[66] A. Jaeger, G. Weiser, P. Wiedemann, I. Gyuro and E. Zielinski, The sizes of coherent band states in semiconductors derived from the Franz-Keldysh effect, Journal of Physics: Condensed Matter **8**, 6779 (1996).

[67] A. Jaeger and G. Weiser, Excitonic electroabsorption spectra and Franz-Keldysh effect of $In_{0.53}Ga_{0.47}AsInP$ studied by small modulation of static fields, Phys. Rev. B **58**, 10674 (1998).

[68] H. J. Kolbe, C. Agert, W. Stolz and G. Weiser, Confinement effects in bulk samples derived from the Franz-Keldysh effect, Phys. Rev. B **59**, 14896 (1999).

[69] H. J. Kolbe, C. Agert, W. Stolz and G. Weiser, Coherence in real space: The transition range from bulk to confined states studied by the Franz-Keldysh effect, Physica E: Low-dimensional Systems and Nanostructures **6**, 173 (2000).

[70] S. Kumar Saha, S. Manna, V. S. Stepanyuk and J. Kirschner, Visualizing non-abrupt transition of quantum well states at stepped silver surfaces, Scientific Reports **5**, 12847 (2015).

# Supplementary Materials for Universal 3:1 Scaling of Quantum-Confined Stark Spectra Revealed by a Three-Dimensional Profile

Sha Han[1,2], Kebei Chen[1,2], Runnan Zhang[1,2], Juemin Yi[1,2], Wentao Song*[1,2] and Ke Xu*[1,2,3,4]

1) Platform for Characterization and Test, Suzhou Institute of Nano-Tech and Nano-Bionics, Chinese Academy of Sciences (CAS), Suzhou 215123, Jiangsu, People’s Republic of China

2) CAS Key Laboratory of Nanophotonic Materials and Devices, Suzhou Institute of Nano-Tech and Nano-Bionics, Suzhou 215123, People’s Republic of China

3) Suzhou Nanowin Science and Technology Co, Ltd., Suzhou, 215123, Jiangsu, China

4) Shenyang National Laboratory for Materials Science, Jiangsu Institute of Advanced Semiconductors, NW-20, Nanopolis Suzhou, 99 Jinji Lake Avenue, Suzhou Industrial Park, Suzhou 215123, Jiangsu, People’s Republic of China

Corresponding authors: wtsong2017@sinano.ac.cn and kxu2006@sinano.ac.cn

## 1. The scaling laws for GaAs

To validate the universality of the derived scaling laws, we apply the theory to GaAs quantum wells. The material parameters for GaAs are as follows: $m_e = 0.06m_0$, $m_{hh} = 0.50m_0$[1], $E_g$ =1.424 eV and breakdown field strength $F$ ≈0.5 MV/cm [2]. To enable a meaningful comparison with the InGaN system, we ensure that the number of bound states in the GaAs quantum well is comparable to that in the InGaN reference case. The number of bound states is given by $N_i = \frac{2}{3\pi}\frac{\sqrt{2m_i eFL^3}}{\hbar}$ (i for e or h) from $E_{k,i} = C_k E_{0,i} = eFL$. To equate the number of electron bound states in GaAs ($N_e^{\text{GaAs}}$) with that in InGaN ($N_e^{\text{InGaN}}$), we substitute the known parameters:

- $m_e^{\text{InGaN}} = 0.18\ m_0,\ m_e^{\text{GaAs}} = 0.06\ m_0$
- $m_{hh}^{\text{InGaN}} = 2.00\ m_0,\ m_{hh}^{\text{GaAs}} = 0.50\ m_0$
- $F_{InGaN} = 2.0\ \text{MV/cm},\ F_{GaAs} = 0.5\ \text{MV/cm}$

We obtain that $L_{GaAs} = L_{InGAN} \times \left(\frac{m_i^{InGaN}}{m_i^{GaAs}} \times \frac{F_{InGaN}}{F_{GaAs}}\right)^{\frac{1}{3}}$. For electron, $L_{GaAs} = 5.0 \times \left(\frac{1.8}{0.6} \times \frac{2.0}{0.5}\right)^{\frac{1}{3}} \approx 11.4$ nm. For hole, $L_{GaAs} = 5.0 \times \left(\frac{2.0}{0.5} \times \frac{2.0}{0.5}\right)^{\frac{1}{3}} \approx 12.5$ nm. Thus, the minimum GaAs quantum well width corresponding to $L = 5.0$ nm in InGaN is 11 nm. However, 11 nm is too large for the quantum well. Thus, to cover a reasonable range, we simulate $L$ from 7 nm to 14 nm. As shown in Fig. S1, at $F = 0.5$ MV/cm, the average peak spacings in both the sub-bandgap region (blue) and the above-bandgap region (red) for GaAs quantum wells follow the $1/L^2$ scaling. The theoretical prediction from the 3D profile shows the coefficients of 15.218 eV·nm² for $\overline{\Delta E_{tail}}$ and 5.073 eV·nm² for $\overline{\Delta E_{edge}}$. The larger coefficients compared to InGaN are due to the smaller effective masses in GaAs, demonstrating the generality of the derived scaling laws beyond the InGaN system.

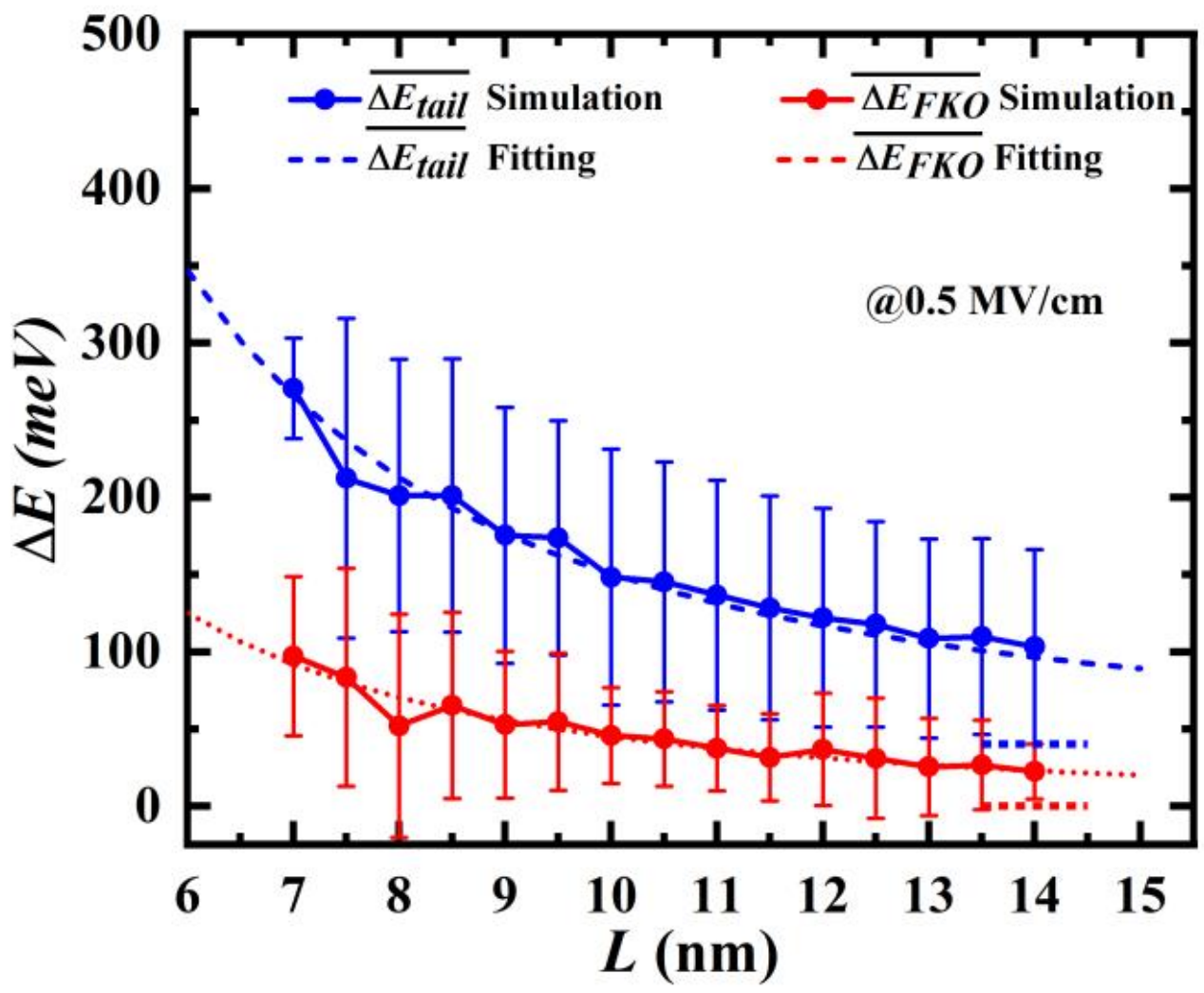

Fig. S1. Universal scaling laws confirmed in GaAs quantum wells with $L$ at $F = 0.5$ MV/cm. Solid lines: simulation data. Dashed lines: fits with coefficients of 11.052 eV·nm² (blue) and 4.500 eV·nm² (red).

## 2. Derivation of the scaling laws

### S1. Theoretical foundation and definitions

We model the bound states in a triangular potential well of depth $E = eFL$ using the WKB approximation. The energy levels for carriers of type $i$ (electron or hole) are:

$$E_{k,i} = C_k E_{0,i} \qquad \text{(Eq.1)}$$

where

- $k$ is the quantum number index.
- $C_k = \left[\frac{3\pi}{2}(k - 0.25)\right]^{2/3}$ is the $k$-th root of the Airy function.
- $E_{0,i} = \left(\frac{e^2F^2\hbar^2}{2m_i}\right)^{1/3}$ is the characteristic energy scale.
- $e$ is the elementary charge, $F$ is the electric field strength, $\hbar$ is the reduced Planck's constant, and $m_i$ is the effective mass ($m_e$ for electrons, $m_h$ for holes).

### S2. The 3D profile and intersection points

In our 3D profile (Fig. 3, main text), the transition energy is $E_e + E_h$. The bandgap line $E_g$ and the first Franz-Keldysh oscillation (FKO) line $E_g + 2.05E_0$ in this profile satisfy:

- **Bandgap line**: $E_e + E_h = E_g$.
- **FKO line**: $E_e + E_h = E_g + 2.05E_0$.

Here, $E_0 = \left(\frac{e^2\hbar^2F^2}{2\mu}\right)^{\frac{1}{3}}$ is the characteristic FKO energy, and $\mu$ is the reduced mass, defined by $\mu^{-1} = m_e^{-1} + m_h^{-1}$.

For a specific electron state $k$, its energy $E_{k,e}$ is fixed. We assume a quasi-continuous spectrum of electron states. The intersections of the vertical line at $E_{k,e}$ with the bandgap line $E_g$ and the FKO line $E_g + 2.05E_0$ define specific hole state indices $j_1(k)$ and $j_2(k)$, respectively. These are determined by:

$$eFL = E_{k,e} + E_{j_1,h} = C_k E_{0,e} + C_{j_1} E_{0,h} \quad \text{(Eq.2)}$$

$$eFL + 2.05E_0 = E_{k,e} + E_{j_2,h} = C_k E_{0,e} + C_{j_2} E_{0,h} \quad \text{(Eq.3)}$$

**S3. Deriving the scaling laws**

**S3.1 Sub-bandgap region ($\overline{\Delta E_{tail}}$)**

From Eq. 2, we solve for $C_{j_1}$:

$$C_{j_1} = \frac{eFL - C_k E_{0,e}}{E_{0,h}} \quad \text{(Eq.4)}$$

Using $C_{j_1} = \left[\frac{3\pi}{2}(j_1(k) - 0.25)\right]^{2/3}$, we solve for $j_1(k)$:

$$j_1(k) = 0.25 + \frac{2}{3\pi}\left(\frac{eFL - E_{0,e}C_k}{E_{0,h}}\right)^{3/2} \quad \text{(Eq.5)}$$

Summing over all electron states $k$ up to $N_e$ (where $C_k E_{0,e} = eFL$), and approximating the sum by an integral:

$$\sum_1^{N_e} j_1(k) = 0.25N_e + \frac{2}{3\pi}\left(\frac{eFL - E_{0,e}C_k}{E_{0,h}}\right)^{3/2} \quad \text{(Eq.6)}$$

For large $L$, the first term dominates:

$$\sum_1^{N_e} j_1(k) \approx \sqrt{m_e m_h}\,\frac{eFL^3}{12\pi\hbar^2} \quad \text{(Eq.7)}$$

The energy span is $eFL$, so the average peak spacing is:

$$\overline{\Delta E_{tail}} = \frac{eFL}{\sqrt{m_e m_h}\frac{qFL^3}{12\pi\hbar^2}} = \frac{12\pi\hbar^2}{\sqrt{m_e m_h}L^2} \quad \text{(Eq.8)}$$

**S3.2 Above-bandgap region ($\overline{\Delta E_{FKO}}$)**

The number of above-bandgap peaks for a given hole state $k$ is $j_2(k) - j_1(k) + 1$. Following a similar procedure using Eq. 2 and Eq. 3, and integrating over $k$, we find:

$$\sum_1^{N_e}(j_2 - j_1 + 1) \approx \frac{2^{\frac{2}{3}}\times 2.05}{8\pi}\frac{\sqrt{m_e m_h}}{\mu^{\frac{1}{3}}}\frac{q^{\frac{2}{3}}L^2F^{\frac{2}{3}}}{\hbar^{\frac{4}{3}}} + N_e \quad \text{(Eq.9)}$$

For large $L$, the first term dominates, giving:

$$\sum_1^{N_e}(j_2 - j_1 + 1) \approx \frac{2^{\frac{2}{3}}\times 2.05}{8\pi}\frac{\sqrt{m_e m_h}}{\mu^{\frac{1}{3}}}\frac{q^{\frac{2}{3}}L^2F^{\frac{2}{3}}}{\hbar^{\frac{4}{3}}} \quad \text{(Eq.10)}$$

The energy span of the above-bandgap region is $2.05E_0$. Therefore, the average peak spacing is:

$$\overline{\Delta E_{FKO}} = \frac{2.05E_0}{\frac{2^{\frac{2}{3}}\times 2.05\sqrt{m_e m_h}q^{\frac{2}{3}}L^2F^{\frac{2}{3}}}{8\pi\,\mu^{\frac{1}{3}}\,\hbar^{\frac{4}{3}}}} = \frac{4\pi\hbar^2}{\sqrt{m_e m_h}L^2} \quad \text{(Eq.11)}$$

## 3. The estimation of the origins for deviation of the theoretical scaling laws

Although the primary model captures the overall scaling behavior, we identify several sources of discrepancy and refine the model accordingly. The first origin is the limitation of the WKB approximation. The WKB approximation introduces a slight overestimation in the coefficients $C_k$ compared to exact solutions. As shown in Table.S1 in SM.3, the maximum difference is only 0.761% with respect to the real value which is negligible compared to the observed ~46% deviation for GaN at 2.0 MV/cm and $L$=5 nm (Fig. 2). Thus, the WKB approximation is sufficiently accurate and not the primary source of discrepancy. The second origin is the overestimation of the sub-bandgap energy span as $eFL$. Table S2 in SM.3 compares this approximation to the actual energy span between the bandgap $E_g$ and the first transition $E_{11}$ for various well widths $L$ at $F$=2.0 MV/cm. The real energy span is significantly smaller than the theoretical span $eFL$ (e.g., 83% smaller at $L$=5 nm). However, the observed model deviation (31% for GaN at $L$=5 nm) is less than this discrepancy, indicating that other factors partially compensate for this overestimation. The third origin is the repeat counting of the point $j_1$ since the line $E_g$ intersects with the vertical electron lines for $N_e$ ($N_e = \frac{2}{3\pi}\frac{\sqrt{2m_e eFL^3}}{\hbar}$ as deduced in the SM.2) times. Thus the number of dots is overestimated and the scaling law is underestimated where the $N_e$ term contributes to the total summation $\sum_1^{N_e} j_1$ by $\frac{8\sqrt{2}\hbar}{\sqrt{m_h eFL^3}} = \frac{10}{\sqrt{m_{h,u}F_u L_u^3}}$ with $F_u$ the electric field strength in unit of MV/cm, $L_u$ the quantum well width in unit of nm and $m_{h,u}$ the hole effective mass in unit of $m_0$. Besides, since the total number of unique optical transitions is a physical invariant and should be independent of the summation path in the 3D profile. Summing vertically over electron states $k$ yields the $N_e$ correction. Conversely, summing horizontally over hole states yields an identical dominant $L^3$ term but a constant correction of $N_h$. Since $N_e$ summation already accurately reflects the physical counting of transitions from the hole-state perspective in Fig.4 and $N_h$ weight $\frac{10}{\sqrt{m_{e,u}F_u L_u^3}}$ at 2.0 MV/cm and 5 nm is 1.5 which is unphysical, we won't change $N_e$ to $N_h$.

The excellent match validates that the initial discrepancy stemmed from this nuance of discrete summation related specifically to electron states. The last origin is the omission term $0.25N_e$ in Eq. 6 in SM.2. Similarly, we won't change the $0.25N_e$ to $0.25N_h$. This correction was omitted in the final scaling law (Eq. 7 in SM.2) as it becomes negligible for large $L$. Physically, it represents a systematic undercount inherent to the integral approximation for a finite number of states. As shown in Figure S2(a), the omission (blue dashed) influences the result slightly. However, the inclusion of the real energy span greatly reduces the $\overline{\Delta E_{tail}}$ (orange dashed) and the inclusion of the repeat counting greatly increases the $\overline{\Delta E_{tail}}$ (pink dashed), where the approximation from the energy span and the repeat counting cancel each other. Taking these 4 factors would improve the final $\overline{\Delta E_{tail}}$ (black dotted).

| **Index k** | **Exact $C_k$** | **WKB's $C_k$** | **Difference (%)** |
|---|---|---|---|
| **1** | 2.3381 | 2.3203 | 0.761% |
| **2** | 4.0879 | 4.0818 | 0.149% |
| **3** | 5.5206 | 5.5172 | 0.062% |
| **4** | 6.7867 | 6.7845 | 0.032% |
| **5** | 7.9441 | 7.9425 | 0.020% |
| **6** | 9.0227 | 9.0214 | 0.014% |
| **7** | 10.0402 | 10.0391 | 0.011% |
| **8** | 11.0085 | 11.0077 | 0.007% |
| **9** | 11.9360 | 11.9353 | 0.006% |
| **10** | 12.8288 | 12.8281 | 0.005% |
| **11** | 13.6915 | 13.6909 | 0.004% |
| **12** | 14.5278 | 14.5273 | 0.003% |
| **13** | 15.3408 | 15.3403 | 0.003% |
| **14** | 16.1327 | 16.1323 | 0.003% |
| **15** | 16.9056 | 16.9053 | 0.002% |
| **16** | 17.6613 | 17.6610 | 0.002% |
| **17** | 18.4011 | 18.4008 | 0.002% |
| **18** | 19.1264 | 19.1261 | 0.002% |
| **19** | 19.8381 | 19.8379 | 0.001% |
| **20** | 20.5373 | 20.5371 | 0.001% |

Table S1. Comparison of the first 20 exact and WKB deduced values of $C_k$ and the difference ratio with respect to the exact $C_k$

| $L$ (nm) | Real Energy Span (eV) | $eFL$ (eV) | Difference (%) |
|---|---|---|---|
| **5** | 0.17 | 1.00 | 83.00% |
| **6** | 0.37 | 1.20 | 69.17% |
| **7** | 0.57 | 1.40 | 59.29% |
| **8** | 0.77 | 1.60 | 51.88% |
| **9** | 0.97 | 1.80 | 46.11% |
| **10** | 1.17 | 2.00 | 41.50% |
| **11** | 1.37 | 2.20 | 37.73% |
| **12** | 1.57 | 2.40 | 34.58% |

Table S2. Comparison of the real energy span and the approximated value ($eFL$) for 2.0 MV/cm from 5 nm to 12 nm.

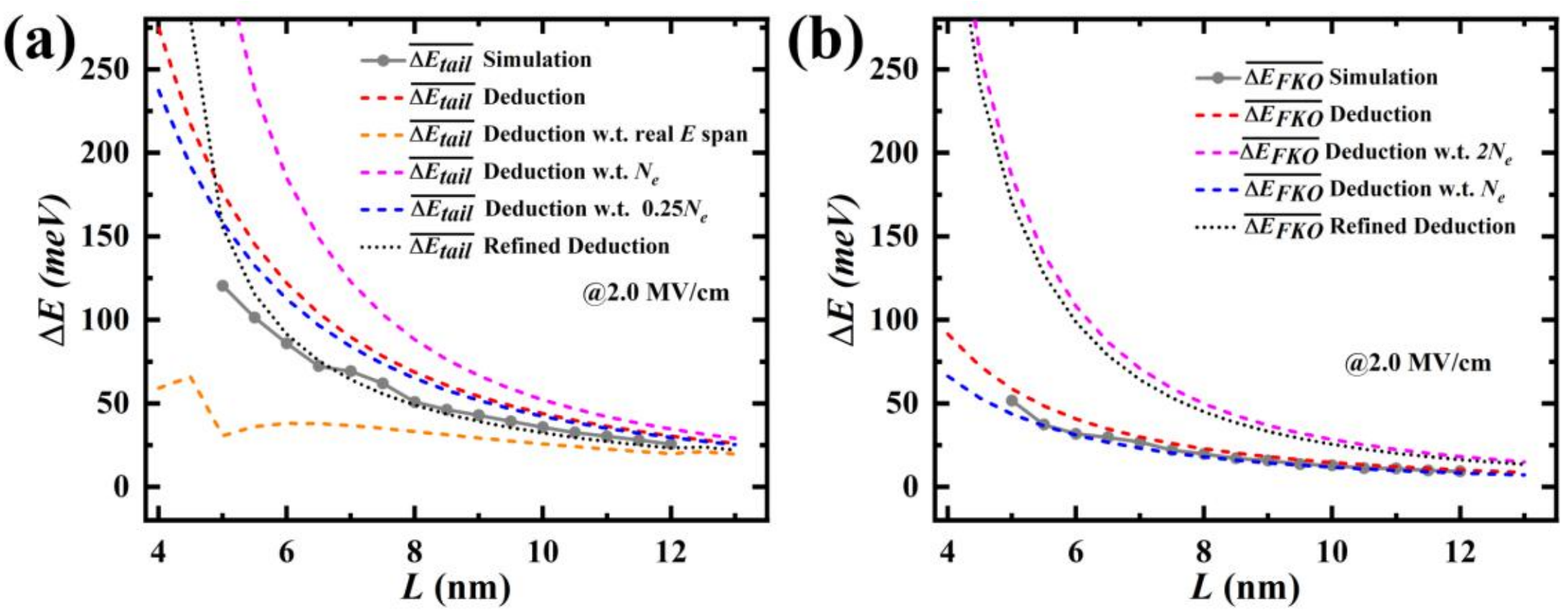

Fig. S2. Refining the scaling laws. Simulation data (gray scatters) for (a) $\overline{\Delta E_{tail}}$ and (b) $\overline{\Delta E_{FKO}}$ are compared against the primary theoretical model (dotted lines, from Eq. 8 & 11 in SM.2) and the refined model (dashed lines). The refined model incorporates all the 3 corrections leads to improved agreement (black dotted) compared to the theoretical results (red dashed) for $\overline{\Delta E_{tail}}$. The corrections for $\overline{\Delta E_{tail}}$ include the real energy span (orange dashed), the repeat counting (pink dashed) and the omission (blue dashed). For $\overline{\Delta E_{FKO}}$, the refined model including the repeat counting (pink dashed) and the omission (blue dashed) results in a larger value (black dotted) than deductions (red).

As shown in Figure S2(b), incorporating a similar $2N_e$ term from repeat counting leads to a large overestimation compared to simulations (pink dashed) and a discrete correction $N_e$ term for the omission leads to an underestimation compared to simulations (blue dashed) where the $N_e$ over total number $\sum_1^{N_e}(j_2 - j_1 + 1)$ is $\frac{16\sqrt{2}\hbar^{\frac{1}{3}}}{3\times 2.05\times 2^{\frac{2}{3}}}\frac{\mu^{\frac{1}{3}}}{\sqrt{m_h}}\frac{1}{(eFL^3)^{\frac{1}{6}}} = \frac{2.2124\mu_u^{\frac{1}{3}}}{\sqrt{m_{h,u}}(F_u L_u^3)^{\frac{1}{6}}}$ with $\mu_u$ the effective reduced mass. This suggests that the original approximation (Eq. 11 in SM.2) is adequate. Taking these 2 factors would improve the final $\overline{\Delta E_{FKO}}$ to some extent (black dotted).

The error bars in Figure 2(b), representing deviations from the average peak spacing, originate from distinct physical reasons in the sub-bandgap and above-bandgap regions. In the sub-bandgap region, the standard deviation arises from the non-uniform distribution of energy levels. As the quantum number $k$ increases, the energy $E_k$ saturates [3] (Table S1 in SM.3). Consequently, the state densities along the electron energy ($E_e$) or hole energy ($E_h$) axes transition from a sparse to a condensed distribution (Fig.S2 in SM.4). This leads to spectral peaks that become increasingly dense near the bandgap energy $E_g$, resulting in significantly smaller energy spacings in that vicinity. The identification of these closely spaced peaks as distinct features causes the standard deviation estimation to approach a large value. In contrast, the standard deviations in the above-bandgap region are considerably smaller. This region corresponds to the transition zone between the Airy-function-like solutions (associated with the triangular potential well of depth $eFL$) and the sine-function-like solutions above the well. Within this transition region, the energy peaks are more uniformly distributed (Fig.S2 in SM.4), leading to reduced uncertainty in the average spacing.

4. **The estimation for the error bars**

Fig. S2 illustrates the distribution of the energy spacings for the sub-bandgap region (blue) and above-bandgap (red) region increasing the energy index for $F$= 2.0 MV/cm and $L$=12 nm. The $\Delta E_{tail}$ decreases significantly from lower energy to higher energy while the $\Delta E_{FKO}$ are very uniform.

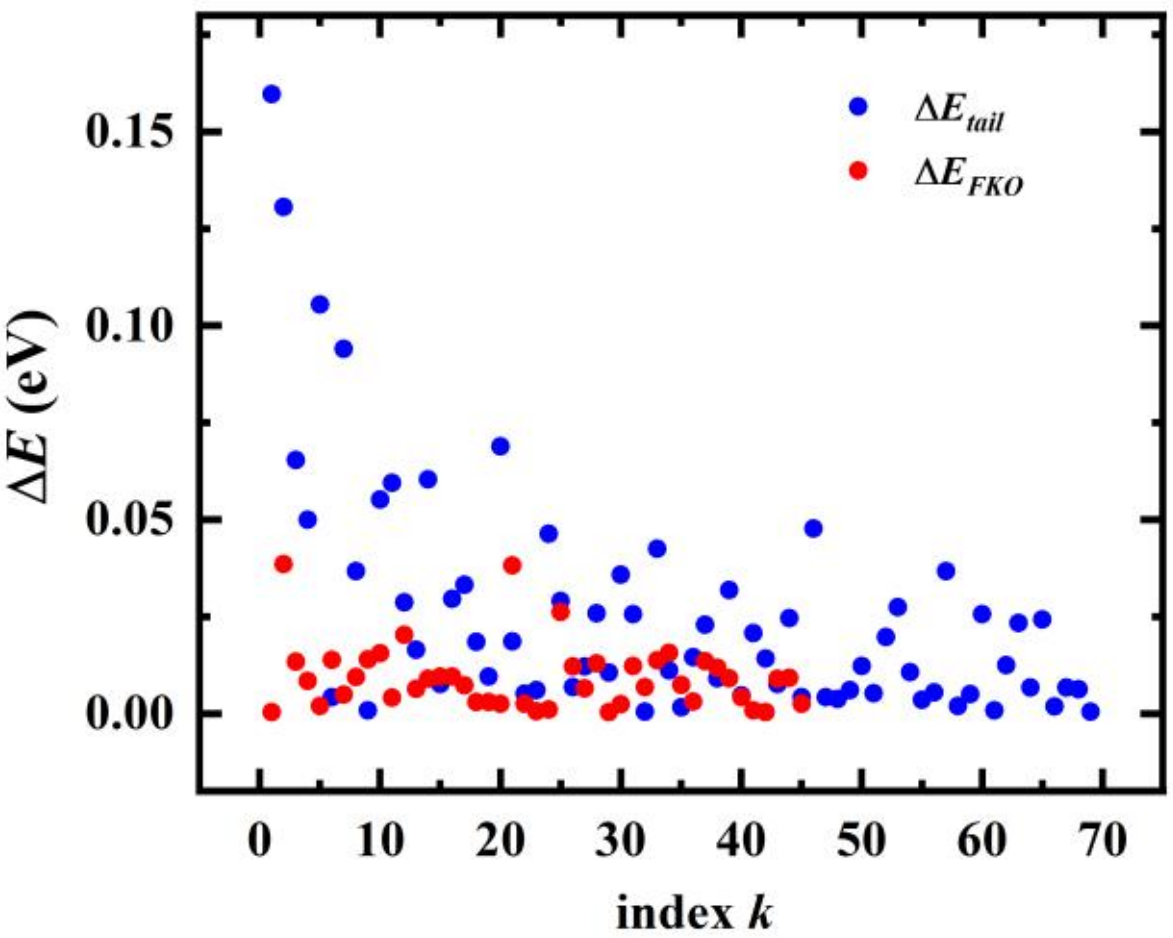

Fig. S2. The distribution of the energy spacings for the sub-bandgap region (blue) and above-bandgap (red) region increasing the energy index for $F$= 2.0 MV/cm and $L$=12 nm.

5. **The typical 3D profile for zero-field, uniform field and non-uniform field**

Fig.S3 compares the 3D profiles and corresponding absorption spectra under three distinct field configurations. In the zero-field case (a1, a2), the band edges are flat, resulting in a uniform grid of allowed transitions in the 3D profile (a1). The absorption spectrum (a2) exhibits a characteristic Tauc background ($\alpha \propto n \propto \sqrt{E}$), reflecting the density of states of a square quantum well. The red dashed line in (a1) with slope $\frac{m_e}{m_h}$ traces this background. Besides, non-zero though small transitions appear below $E_g$ (blue dots under the black dashed line) since the wavefunctions under the electric field is the linear combinations of the eigen wavefunctions ($\sum a_i \Phi_i$) in the square quantum well and $a_i a_j$ constitute the nonzero transition strength. Under a uniform field (b1, b2), band tilting introduces Franz-Keldysh oscillations (FKO). The 3D profile (b1) reveals distinct 135° stripes, corresponding to constant transition energy lines. The minima between these stripes align with the FKO extrema observed in the absorption spectrum (b2). For a non-uniform field (c1, c2), potential disorder disrupts the phase coherence essential for FKO formation. Consequently, the 135° features in the 3D profile (c1) become scattered, and the corresponding absorption spectrum (c2) lacks clear oscillatory behavior.

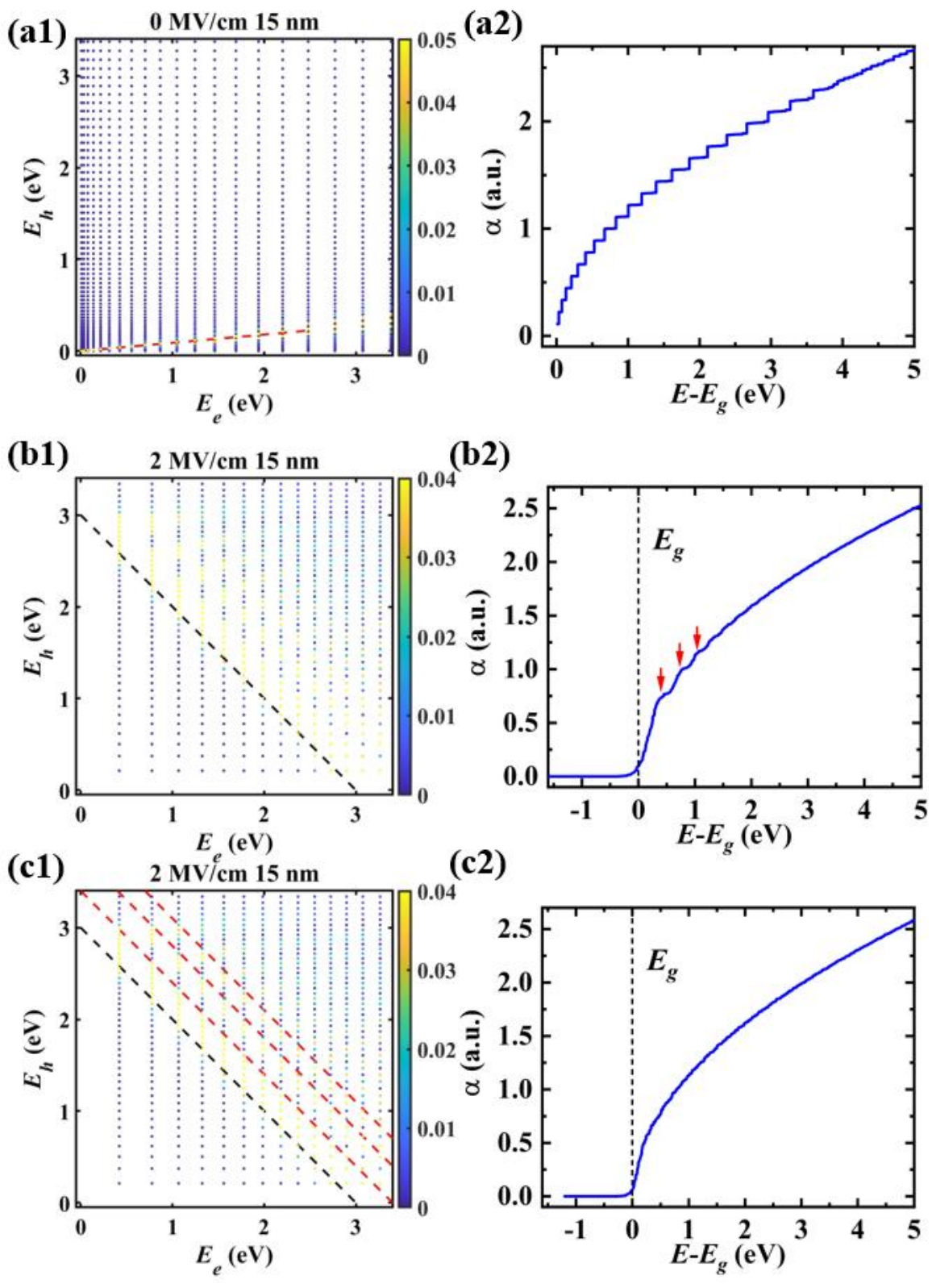

Fig. S3. The 3D profile and absorption spectra across field configurations. (a1, a2) Zero Field: The 3D profile (a1) shows a uniform grid of allowed transitions. In the absence of an electric field, the band edges are flat, eliminating the quantum interference that gives rise to Franz-Keldysh oscillations (FKO). Consequently, the absorption spectrum (a2) exhibits a characteristic Tauc background (, governed solely by the density of states in the square quantum well. The red dashed line in (a1) has slope $m_e/m_h$. (b1, b2) Uniform Field: Applying a uniform field tilts the bands, enabling the FKO. Distinct 135° stripes emerge in the 3D profile (b1), corresponding to these oscillations. Minima between these stripes (below yellow markers) align with FKO extrema in the absorption spectrum (b2, red arrows). (c1, c2) Non-uniform Field: A disordered potential disrupts the phase coherence necessary for well-defined FKO. The 3D profile (c1) shows scattered 135° features, and the absorption (c2) lacks clear oscillatory behavior, highlighting the loss of coherence under non-uniform fields.

### 6. Geometric relationship between quantum well width and coherence length

The 3D profile offers an intuitive explanation for the experimentally observed similarity between the quantum well width $L$ and the coherence length $L_{ch}$ associated with FKO. Our 3D visualization clarifies the origin. As shown in Fig. S3, high-energy states in the profile resemble those of a square quantum well and contribute to Tauc-background-like features. The segment CE in the schematic defines the upper limit for these Tauc-background-like features. From this geometry, we deduce the relation: $\frac{L_{ch}}{L} < \frac{BE}{OA} = \frac{1}{\sqrt{2}}\frac{m_h+m_e}{m_h-m_e}$. This ratio provides an upper bound for $\frac{L_{ch}}{L}$, naturally explaining their observed similarity.

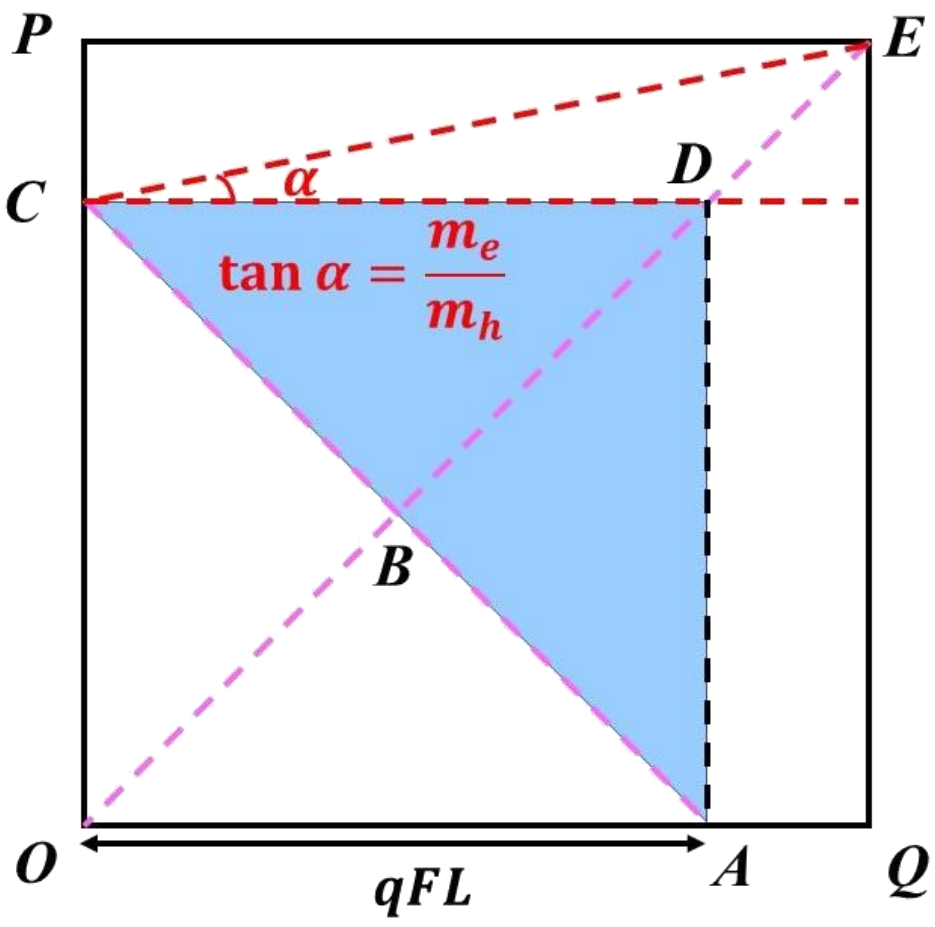

Fig. S4. Schematic link between quantum well width $L$ and coherence length $L_{ch}$. The segment CE defines the upper limit for Tauc-background-like features. The ratio $BE/OA$ provides an upper bound for $L_{ch}/L$, offering a natural explanation for their observed similarity [51-54].

**References**

[1] N. Bouarissa and H. Aourag, Effective masses of electrons and heavy holes in InAs, InSb, GaSb, GaAs and some of their ternary compounds, Infrared physics & techonology **40**, 343 (1999).

[2] S. Tautz, S. Opel, P. Kiesel, S. U. Dankowski, H. M. Hauenstein, A. Seilmeier, M. Krause, U. D. Keif and G. H. Döhler, Spectrally and temporally resolved investigations of the carrier dynamics in biased low temperature grown GaAs, in *European Quantum Electronics Conferenc* (Optica Publishing Group, 1998) p. QMC4.

[3] J. H. Davises, *The physics of low-dimensional semiconductor: an introduction* (Cambridge university press, 1998).